\documentclass[aps,prb,twocolumn,groupedaddress,floatfix,showpacs,amsmath,amssymb]{revtex4}   
\usepackage{graphicx}
\usepackage{amsmath}
\usepackage{psfrag}
\begin{document}
\newcommand\ket[1]{|#1\rangle}
\newcommand\bra[1]{\langle#1|}
\newcommand\braket[2]{\left\langle#1\left|#2\right.\right\rangle}

\title{Charge and spin dynamics in interacting quantum dots} 

\author{Janine Splettstoesser$^{1,2}$, Michele Governale$^{3}$, J\"urgen K\"onig$^{4}$, and Markus B\"uttiker$^{5}$}
\affiliation{$^{1}$ Institut f\"ur Theoretische Physik A, RWTH Aachen University, D-52056 Aachen, Germany\\
$^{2}$ JARA - Fundamentals of Future Information Technology, Germany\\
$^{3}$ School of Physical and Chemical Sciences and MacDiarmid Institute for Advanced Materials and Nanotechnology, Victoria University of Wellington, PO Box 600, Wellington 6140, New Zealand\\
$^{4}$ Theoretische Physik, Universit\"at Duisburg-Essen and CeNIDE, D-47048 Duisburg, Germany\\
$^{5}$D\'epartement de Physique Th\'eorique, Universit\'e de Gen\`eve,  CH-1211 Gen\`eve 4, Switzerland}
\date{\today}

\begin{abstract}
The transient response of a quantum dot with strong Coulomb interaction to a fast change in the gate potential, as well as the stationary ac-response to a slow harmonic variation in the gate potential are computed by means of a real-time diagrammatic expansion in the tunnel-coupling strength.
We find that after a fast switching, the exponential relaxation behavior of charge and spin are governed by a single time constant each, which differ from each other due to Coulomb repulsion.  
We compare the response to a step potential with the $RC$ time extracted from the ac response. 
\end{abstract}
\pacs{73.23.-b,73.23.Hk}
\maketitle

\section{Introduction}
\begin{figure}[b]
\includegraphics[width=3.in]{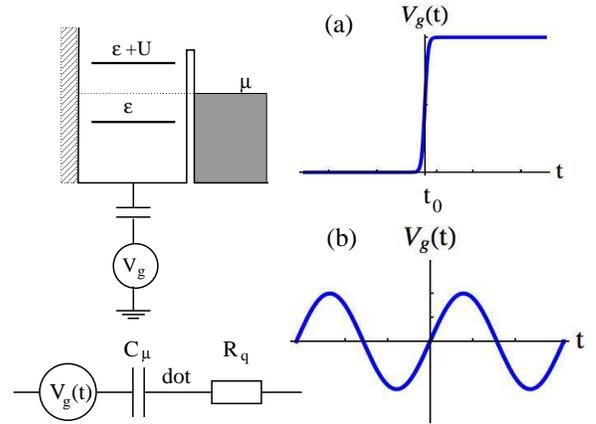}
\caption{Model for the single-level quantum dot, with onsite Coulomb interaction $U$, and its circuit equivalent. (a) A step potential is applied to the dot at time $t=t_0$. (b) The dot is modulated by a slow harmonic potential, which for the extraction of the $RC$-time we assume to be small.  }
\label{fig_model}
\end{figure}

Single electron charging of nanoscale devices driven by a time-dependent signal provides 
novel electron sources for the investigation of mesoscopic transport.~\cite{feve07,blumenthal07} 
Understanding their charging behavior is 
of crucial importance to assess maximum operation speed. 
From the theoretical perspective such an investigation can shed light on the influence of 
quantum-mechanical processes on the time-dependent behavior of nanoscale systems. 
In this work, we focus on the charging 
of a quantum dot, which acts as a mesoscopic capacitor plate, the other plate is a macroscopic gate.~\cite{buttiker93,gabelli06}
This system, when subject to dynamical modulations, 
is the key ingredient to realize a coherent single-electron source, as shown both experimentally~\cite{feve07} and theoretically.~\cite{moskalets08}
The previous studies of mesoscopic capacitors showed that the linear low-frequency response regime is characterized by a quantum electrochemical capacitance and a  charge relaxation resistance, cf. the circuit in Fig.~\ref{fig_model}. When the capacitor is attached to a single-channel lead, the charge relaxation resistance in the fully coherent regime is found to be quantized~\cite{buttiker93,gabelli06,nigg06,wang07}
demonstrating quantum mechanical non-locality of impedances. Theoretical discussions have addressed the regime of weak Coulomb interaction taken into account on the Hartree-Fock level~\cite{nigg06} 
or more recently in the Luttinger-liquid regime, which can be treated by bosonization methods.~\cite{martin09}

In this paper,  we study the time response of a single-level quantum dot driven out of equilibrium by either the fast switching (step pulse) of the gate potential or by a slow periodic modulation.\\
The charge relaxation time after a fast switching has been experimentally extracted in, e.g., Ref. \onlinecite{feve07}.
Also the time scale of spin relaxation in a single quantum dot was measured,~\cite{spinexp} where the relaxation of the spin is due to spin-orbit or hyperfine interaction. For theoretical investigations of these subjects, see, e.g., Refs. \onlinecite{spintheory}. Spin relaxation due to tunneling through a double dot in the spin blockade regime was analyzed in Ref. \onlinecite{vorontsov09}.
Here we are interested in the charge as well as spin relaxation in an interacting quantum dot, uniquely due to electrons tunneling to the reservoirs.

While Coulomb interaction is weak in some of the experiments,~\cite{feve07,gabelli06} side-gate coupled semiconductor quantum dots~\cite{blumenthal07} 
often exhibit strong local Coulomb interaction effects and, therefore, a theory applicable to this regime is highly desirable. 
We consider a quantum dot with arbitrarily strong onsite Coulomb interaction, coupled weakly to a reservoir by a tunneling contact. 
We use a diagrammatic real-time approach to study the relaxation behavior of such a mesoscopic capacitor beyond the Markov limit, taking into account tunneling processes up to next-to-leading order in the tunnel coupling (i.e. fourth order in the tunnel matrix elements). 
This expansion permits arbitrary interaction strength $U$. Both $U$ and temperature are assumed to be large compared to the line width $\Gamma$ of the single level state. 
The effect of Coulomb interaction is manifest  already in lowest order in the tunnel coupling, introducing a dependence on the level position of the relaxation times. We find that charge and spin have \textit{independent} dynamics with a charge relaxation time that is reduced and spin relaxation time that is enhanced by interaction. 
Higher-order tunneling contributions lead to a difference between the $RC$ time extracted from the ac admittance and the relaxation time extracted from the exponential decay after a fast  switch. This is in contrast to the behavior of a classical $RC$ circuit and signalizes that care must be taken in extracting the characteristic time-scales of the system. 

\section{Model and formalism}
We consider a quantum dot attached to a single lead by a tunneling contact, cf. Fig.~\ref{fig_model}, described by the Hamiltonian $H=H_\mathrm{dot}+H_\mathrm{tunnel}+H_\mathrm{lead}$. The single-particle level spacing in the dot is assumed to be larger than any other energy scale (temperature, Coulomb interaction, gate voltage and the highest frequency in the time-dependent signals) such that only one energy level is accessible. The dot can be  described by the single-level Anderson model 
\begin{equation}
H_\mathrm{dot}=\sum_{\sigma=\uparrow,\downarrow}\epsilon(t) d^{\dagger}_\sigma d^{}_\sigma
+U n_\uparrow n_\downarrow, 
\end{equation}
with $d^{\dagger}_\sigma(d^{}_\sigma)$
 being the creation (annihilation) operator for an electron  with spin $\sigma$ on the dot,  and $n_\sigma=d^\dagger_\sigma d_\sigma$ the corresponding number operator. The level position is assumed to depend on time via a gate potential. The onsite repulsion  $U$ describes the energy cost for double occupation and stems from Coulomb interaction (within the constant interaction model~\cite{constant_int}).
Tunneling of electrons between the dot and  the lead is taken into account  by 
$H_\mathrm{tunnel} =\sum_{k,\sigma} V c^{\dagger}_{k,\sigma}d^{}_\sigma +\mathrm{h.c.}$, 
where we assume a momentum-independent tunnel matrix element  $V$ and define the creation (annihilation) operators $c^{\dagger}_{k,\sigma}(c^{}_{k,\sigma})$ for electrons with spin $\sigma$ and momentum $k$ in the lead. The lead Hamiltonian is given by $H_\mathrm{lead}=\sum_{k,\sigma}\epsilon_k c^{\dagger}_{k,\sigma} c^{}_{k,\sigma}$.
We assume a constant density of states $\rho$ in the leads and define the tunnel coupling strength $\Gamma$ as $\Gamma=2\pi\rho|V|^2$.  \\
Since we are not interested in the dynamics of the non-interacting leads' degrees of freedom, we trace them out obtaining an effective description of the dot in terms of its reduced density matrix. We choose as basis of the four-dimensional Hilbert space of the reduced system the eigenstates, $|\chi\rangle$, of the isolated dot Hamiltonian $H_{\text{dot}}$: $|0\rangle$ for an empty dot, $|\sigma\rangle$ for a singly occupied dot
with spin $\sigma=\uparrow,\downarrow$
and $|d\rangle$ for a doubly occupied dot.
Assuming spin-independent  tunneling, we can restrict our interest to the time
evolution of the diagonal elements of the reduced density matrix, $p_\chi$. 
These probabilities are arranged in the vector $\mathbf{P}=(p_{0},p_{\uparrow},p_{\downarrow},p_{\mathrm{d}})$, whose time evolution is given by the generalized master equation
\begin{eqnarray}\label{eq_master}
	\frac{d\mathbf{P}(t)}{dt}=\int_{t_0}^{t}dt'\mathbf{W}(t,t')\mathbf{P}(t')\ .
\end{eqnarray}
The elements $W_{\chi,\chi'}(t,t')$ of the matrix
$\mathbf{W}(t,t')$ describe transitions from state $\chi'$ at time $t'$ 
to state $\chi$ at time $t$.

\section{Fast switching}
We first consider the time evolution after a fast switching event at $t=t_0$, as shown in Fig.~\ref{fig_model} (a).
The amplitude of the switching is only limited insofar as we assume that only the single level of the dot is accessible, i.e. the amplitude is smaller than the level spacing.
The initial value, i.e., the equilibrium distribution before changing the gate voltage, is denoted by 
$\mathbf{P}^\mathrm{in}$. 
In order to identify the relaxation time, we perform the following steps.
First, we expand the probability vector $\mathbf{P}(t')$ on the right-hand side of 
Eq.~(\ref{eq_master}) around the final time $t$, i.e. $\mathbf{P}(t')=\sum_{n=0}^\infty\frac{1}{n!}\left(t'-t\right)^n\frac{d^n\mathbf{P}(t')}{dt'^n}|_{t'=t}$.
Second, we realize that for $t'>t_0$, the Hamiltonian is time independent, and hence the transition
matrix elements depend only on the difference of the time arguments, $\mathbf{W}(t,t')\rightarrow\mathbf{W}(t-t')$.
Third, we replace the lower integration bound $t_0$ by $-\infty$.
This is justified for times $t$ that are larger than the characteristic time scale over which the kernel 
$\mathbf{W}(t)$ decays.
The regime of shorter times $t$ is not the subject of the present work.
As a result of these three steps, we obtain
\begin{equation}
\label{eq_master2}
	\frac{d\mathbf{P}(t)}{dt} = 
		\sum_{n=0}^\infty \frac{1}{n!} \partial^n \mathbf{W}\cdot\frac{d^n\mathbf{P}(t)}{dt^n} \, ,
\end{equation}  
introducing the derivatives of the Laplace transforms, 
$\partial^n\mathbf{W}:=\frac{\partial^n}{\partial z^n} \left[ \int_{-\infty}^t dt' \ e^{-z(t-t')}\ \mathbf{W}(t-t')\right]_{z=0^+}$.
The formal solution of Eq.~(\ref{eq_master2}) can be written as 
\begin{equation}\label{eq_formal_sol}
\mathbf{P}(t)  = 
	  \exp\left( \mathbf{A} t\right) \mathbf{P}^\mathrm{in}. 
\end{equation}
The matrix $\mathbf{A}$ always possesses the eigenvalue $0$, which guarantees the existence of a stationary probability distribution. The other eigenvalues of $\mathbf{A}$ define the relaxation rates.
In the following, we concentrate on the limit of weak tunnel coupling between quantum dot and lead.
This motivates an expansion of $\mathbf{A}=\mathbf{A}^{(1)}+\mathbf{A}^{(2)}+\ldots$ in powers of the tunnel-coupling strength $\Gamma$, indicated by the  superscript $(i)$.
Substituting the formal solution for $\mathbf{P}(t)$, Eq. (\ref{eq_formal_sol}),  into Eq.~(\ref{eq_master2}) iteratively yields
$\mathbf{A}^{(1)} = \mathbf{W}^{(1)}$ and $\mathbf{A}^{(2)} = \mathbf{W}^{(2)} +\partial\mathbf{W}^{(1)} \cdot \mathbf{W}^{(1)}$. 
Note that non-Markovian  contributions, i.e. contributions proportional to  $\partial^n\mathbf{W}$, enter only starting from second order in $\Gamma$.  The kernel $\mathbf{W}(z)$ can be computed by the real-time diagrammatic approach developed in 
Ref.~\onlinecite{koenig96}.

We are interested in the inverse relaxation times of charge $\gamma_\mathrm{c}=1/\tau_\mathrm{c}$ and spin  $\gamma_\mathrm{s}=1/\tau_\mathrm{s}$. They are found as the eigenvalues of $\mathbf{A}$ related to the left eigenvectors 
${\bf v}_c =  \left(0,1,1,2\right)-\langle n_c\rangle^{eq}\left(1,1,1,1\right)$ and ${\bf v}_s =  \left(0,1,-1,0\right)$, 
with the equilibrium occupation number $\langle n_c\rangle^\mathrm{eq}$. This yields the expressions for the time-dependent expectation value of the charge $\langle n_c \rangle(t)=
\left(0,1,1,2\right) \cdot \mathbf{P}(t)$ and the spin $\langle n_s \rangle(t)= \left(0,1,-1,0\right) \cdot \mathbf{P}(t)$ on the dot
\begin{eqnarray*}
\langle n_c \rangle(t) & = & \langle n_c\rangle^\mathrm{eq}\left(1-e^{-t/\tau_c}\right)+\langle n_c \rangle^\mathrm{in}e^{-t/\tau_c}\\
\langle n_s \rangle(t) & = & \langle n_s \rangle^\mathrm{in}e^{-t/\tau_s}\ ,
\end{eqnarray*}
with the initial occupation numbers indicated by the superscript "in".
The spin relaxation time can be measured by preparing the dot in a spin-polarized state by applying a magnetic field and 
looking at the spin relaxation after the field has been switched off. 
If we retain only the first-order contribution in $\Gamma$, the inverse relaxation times are given by,~\cite{splett08-1}
\begin{subequations}
\begin{eqnarray}
\gamma_\mathrm{s}^{(1)} & = & \Gamma\left[1-f(\epsilon)+f(\epsilon+U)\right]\\
\gamma_\mathrm{c}^{(1)} & = & \Gamma\left[1+f(\epsilon)-f(\epsilon+U)\right],
\end{eqnarray}
\end{subequations}
with the Fermi functions $f(\omega)$. 
\begin{figure}[t]
\includegraphics[width=3.in]{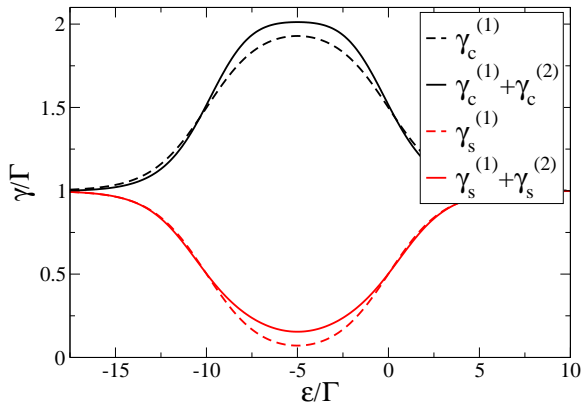}
\caption{Charge relaxation rate, $\gamma_\mathrm{c}$, and spin relaxation rate, $\gamma_\mathrm{s}$, in first and second order in the tunnel coupling $\Gamma$ for $U=10\Gamma$ as a function of the dot's level position $\epsilon$ in units of $\Gamma$, $k_\mathrm{B}T=1.5\Gamma$.}
\label{fig_rel}
\end{figure}
For vanishing Coulomb interaction, $U=0$, the relaxation times are independent of the level position $\epsilon$ and both equal to  the inverse of the tunnel coupling $\Gamma$. 
In  the presence of a finite $U$, when the level position is in the interval  $-U<\epsilon<0$ the dot is predominantly singly occupied.  In this parameter range, $\gamma_\mathrm{c}$  and  $\gamma_\mathrm{s}$ differ strongly, even though in first order in the tunnel coupling a change in the spin occupation  goes always along with a change in the  charge occupation.  
There are two possible states for the dot to relax to single occupation, namely $|\uparrow\rangle, \ |\downarrow\rangle$,  and consequently the charge relaxation time is decreased by a factor $1/2$. On the other hand, in first order in tunneling the spin is blocked to a value different from 
its equilibrium value $0$
due to the lack of spin-flipping processes leading to an enhancement of $\tau_\mathrm{s}$. This is shown in Fig. (\ref{fig_rel}), where the inverse relaxation times for spin, $\gamma_\mathrm{s}$, and charge, $\gamma_\mathrm{c}$, are plotted as a function of the new dot's level position after the switching, for finite $U$. 
Note that spin and charge have independent dynamics; this is not the case in the presence of a Zeeman splitting in the dot.  In fact, in the presence of a Zeeman splitting, 
           $\epsilon_\uparrow\neq\epsilon_  \downarrow$, and for 
infinite $U$, the first order relaxation rates
            are given by $1/\tau_1=1+\sqrt{f(\epsilon_\uparrow)f(\epsilon_\downarrow)}$ and 
            $1/\tau_2=1-\sqrt{f(\epsilon_\uparrow)f(\epsilon_\downarrow)}$.\\
When we include tunneling contributions in second order in the tunnel coupling, we find corrections of different origin to the relaxation times. Cotunneling, describing real tunneling processes via energetically forbidden intermediate states, leads to coherent processes flipping the spin of the dot, $W_\mathrm{sf}$, or to transitions between zero and double occupation, $W_{0\mathrm{d}}$ and $W_{\mathrm{d}0}$. These terms, which agree with results  from  a standard second-order perturbation expansion, contain expressions 
$\phi(\epsilon)=\frac{1}{2\pi}\mathrm{Re}\Psi\left(\frac{1}{2}+i\frac{\beta\epsilon}{2\pi}\right)$ and $\sigma(\epsilon,U)=\Gamma[\phi(\epsilon+U)-\phi(\epsilon)]$, where $\Psi(x)$ is the digamma function and $\beta=1/(k_{\text{B}}T)$ the inverse temperature, primes denote derivatives with respect to $\epsilon$.
 These cotunneling rates are given by
\begin{eqnarray*}
W_\mathrm{sf} & = & -\frac{\Gamma}{\beta}\left[\Gamma\phi''(\epsilon)+\Gamma\phi''(\epsilon+U)-\frac{2}{U}\sigma'(\epsilon,U)\right]\\
W_{\mathrm{d}0} & = & - \frac{2 \Gamma} {e^{\beta(2\epsilon+U)}-1}  \left[\Gamma\phi'(\epsilon)+\Gamma\phi'(\epsilon+U)-\frac{2}{U}\sigma(\epsilon,U)\right],
\end{eqnarray*}
and $W_{0\mathrm{d}} =  \exp[\beta(2\epsilon+U)] W_{\mathrm{d}0}$.
In addition to the cotunneling terms virtual tunneling processes appear, which can be captured in terms of renormalization of the level position, $\epsilon\rightarrow\epsilon+\sigma(\epsilon,U)$, and the tunnel coupling, $\Gamma\rightarrow\Gamma[1+\sigma_\Gamma(\epsilon,U)]$,~\cite{comment_renormalization} with
 \begin{eqnarray}\label{eq_gammaren}
\sigma_\Gamma
(\epsilon,\Gamma,U) & = &  -\left[\Gamma\phi'(\epsilon)+\Gamma\phi'(\epsilon+U)-\frac{2}{U}
                 \sigma(\epsilon,\Gamma,U)
       \right]\nonumber
\\
&& \times
       \frac{1-f(\epsilon)-f(\epsilon+U)}{1-f(\epsilon)+f(\epsilon+U)}\ .
\end{eqnarray}
These corrections are due to an interplay of tunneling and interaction and vanish for $U=0$. 
We verified that they can also be identified in the second-order linear conductance.

The corrections to the 
relaxation times read 
\begin{subequations}
\begin{eqnarray}
\gamma_\mathrm{s}^{(2)} & = & \sigma(\epsilon,U) \frac{\partial}{\partial\epsilon}\gamma_\mathrm{s}^{(1)}+\sigma_\Gamma(\epsilon,U)\gamma_\mathrm{s}^{(1)}+2W_\mathrm{sf}\\
\gamma_\mathrm{c}^{(2)} & = &  \sigma(\epsilon,U)\frac{\partial}{\partial\epsilon}\gamma_\mathrm{c}^{(1)}+ \sigma_\Gamma(\epsilon,U)\gamma_\mathrm{c}^{(1)}\nonumber\\
 & &+2 \frac{
                 f(\epsilon+U)W_{0\mathrm{d}}
                 +
                  \left(1-f(\epsilon)\right)W_{\mathrm{d}0}
                  }
                   {1-f(\epsilon)+f(\epsilon+U)} \ .
\end{eqnarray}
\end{subequations}
The influence of the corrections on the relaxation times is shown in Fig.~\ref{fig_rel}. Both relaxation rates are enhanced due to the higher order tunneling processes. Importantly, the charge and spin relaxation dynamics remain governed by a single time 
scale, each.

\section{Harmonic modulation}
We now consider a harmonic modulation of the level position $\epsilon$ with frequency $\Omega$, which we assume to be much smaller than $\Gamma$, cf. Fig.~\ref{fig_model}(b). 
We perform a systematic expansion in powers of $\Omega$.~\cite{splett06,cavaliere09}
The order in the expansion in $\Omega$, is indicated by a superscript: $(\mathrm{i})$ (instantaneous) for  zeroth order and $(\mathrm{a})$ (adiabatic) for first order.
While the instantaneous term for the average occupation of the dot, $\langle n \rangle^{(i)}$, immediately adjusts to the applied gate voltage, its first-order correction, $\langle n \rangle^{(a)}$ accounts for  a slight lagging behind.

In the ac regime, the quantum dot can be modeled by its charge relaxation resistance, $R_q$, in series with its electrochemical capacitance, $C_\mu$.~\cite{buttiker93}
We are interested in the frequency-dependent admittance, $G(\Omega)$, which is defined by the linear relation $G(\Omega)=I(\Omega)V(\Omega)$. Therefore we consider the linear regime in the gate potential controlling the level position $\epsilon$. 
The expansion of the admittance of the dot 
\begin{equation}
G(\Omega) = -i\Omega C_\mu+\Omega^2C_\mu^2R_q \ ,
\end{equation}
 for small frequencies enables to identify these quantities and to compute the $RC$ time. 
To compare with this, we  insert the adiabatic expansion of the occupation number  into the charge continuity equation $I(t) = -e\ \partial\langle n\rangle/\partial t$ (Ref. \onlinecite{note_displacement}) to arrive at
\begin{equation}
	\frac{1}{\tau_{RC}} = -\frac{\partial \langle n \rangle^{(i)}/\partial t}{\langle n \rangle^{(a)}} \, .
\end{equation}
To obtain the average dot occupation, we expand the master equation in powers of $\Omega$, 
\begin{subequations} 
\label{eq_master_ad}
\begin{eqnarray}
	0 & = & \mathbf{W}^{(i)}\mathbf{P}^{(i)}
\\
	\frac{d \mathbf{P}^{(i)}}{dt} & = & 
		\mathbf{W}^{(i)}\mathbf{P}^{(a)}+\mathbf{W}^{(a)}\mathbf{P}^{(i)}+
			\partial\mathbf{W}^{(i)}\frac{d \mathbf{P}^{(i)}}{dt}\ .
\end{eqnarray}
\end{subequations}
We refer to Ref.~\onlinecite{splett06} for the evaluation of the adiabatic correction of the kernel, $\mathbf{W}^{(a)}$. The instantaneous occupation probabilities, $\mathbf{P}^{(i)}$, and their first adiabatic correction, $\mathbf{P}^{(a)}$, are obtained from the master equation Eq.~(\ref{eq_master_ad}) 
by matching orders  in the tunnel coupling. We find that  the instantaneous probabilities $ \mathbf{P}^{(i)}$ start in zeroth order in the tunnel coupling, while $\mathbf{P}^{(a)}$ starts in minus first order in the tunnel coupling, with $\mathbf{P}^{(a,-1)}$  being proportional to the small factor  $\Omega/\Gamma$.

A rigorous expansion in the tunnel coupling exists for the inverse of the $RC$-time, 
$\gamma_{RC}=1/\tau_{RC}$.
Close to the resonances, $\epsilon=0$ and $\epsilon=-U$, both the change in the instantaneous occupation and the adiabatic correction are governed by lowest order in the tunnel coupling, which leads to the same relaxation time as found for the charge relaxation time due to fast switching,
$\gamma_{RC}^{(1)} =\gamma_{c}^{(1)}\def\gamma^{(1)}$. 
In the Coulomb-blockaded regions, however, both $\partial \langle n \rangle^{(i,0)}/\partial t$ and $\langle n \rangle^{(a,-1)}$ are exponentially suppressed, and the expression for $\gamma_{RC}^{(1)}$ becomes meaningless.
Instead, the next-order corrections in $\Gamma$ needs to be taken into account.
For the time variation in the instantaneous occupation number $\partial \langle n \rangle^{(i,1)}/\partial t$, the non-exponentially suppressed contribution to this  higher-order correction is due to quantum fluctuations, namely cotunneling processes in which the final dot state is not changed as compared to the initial one.
Their rates are $W_{0\rightarrow\sigma\rightarrow 0}=W_{\sigma\rightarrow 0\rightarrow \sigma}=-\frac{\Gamma^2}{\beta}\phi''(\epsilon)$ and $W_{\sigma\rightarrow d\rightarrow \sigma}=W_{d\rightarrow \sigma\rightarrow d}=-\frac{\Gamma^2}{\beta}\phi''(\epsilon+U)$.
The higher-order correction $\langle n \rangle^{(a,0)}$, on the other hand, remains exponentially suppressed. As a result, the relaxation rate diverges. 
This type of processes  presumably contributes only to the short time dynamics after a step pulse and does not enter the long-time behavior described by the exponential  relaxation.  

\begin{figure}[t]
 \includegraphics[width=3.in]{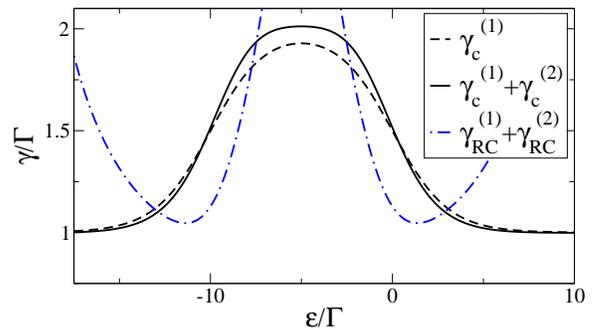}
\caption{Inverse $RC$ time, $\gamma_\mathrm{RC}$, and charge relaxation rate, $\gamma_\mathrm{c}$, up to second order in the tunnel coupling $\Gamma$ for $U=10\Gamma$ as a function of the dot's level position $\epsilon$ in units of $\Gamma$, $k_\mathrm{B}T=1.5\Gamma$.}
\label{fig_compare_times}
 \end{figure}

We find for the difference between the $RC$ rate $\gamma_{RC}$  and the charge relaxation rate $\gamma_{c}$  
 \begin{eqnarray}
	\frac{\gamma_{RC}^{(2)}-\gamma_{c}^{(2)}}{\gamma^{(1)}} &=& \Gamma \frac{\left[2-\langle n\rangle^{(i,0)}	\right]\phi''(\epsilon)+\langle n\rangle^{(i,0)}\phi''(\epsilon+U)}{\partial\langle n\rangle^{(i,0)}/\partial\epsilon}\ .\nonumber\\
 \end{eqnarray} 
 Although this difference is only due to second-order rates, it is substantial in the Coulomb-blockade regions where first-order processes are exponentially suppressed. 
This is shown in Fig.~\ref{fig_compare_times}, where the inverse $RC$-time and the relaxation rate after a single switching event are compared, 
taking into account tunneling processes up to second order. This difference is a consequence of the different time scales captured in the two expansions. Whereas in the response to the step we take into account the long-time limit but omit high frequencies, the ac response is sensitive only to the frequency at which it is tested. 
When taking $U=0$ this result agrees well with the respective limit of Ref.~\onlinecite{buttiker93} in the weak-coupling regime.

\section{Conclusion}
We calculated the relaxation times of a single-level interacting quantum dot contacted to a lead and driven out of equilibrium by a step-pulse potential and the $RC$-time extracted from the ac response. In the presence of Coulomb interaction the relaxation times depend on the position of the dot level, leading to different relaxation times for charge and spin. Corrections in second order of the tunnel coupling contain real cotunneling processes and renormalization corrections to the level position and the coupling strength. 
Cotunneling processes leaving the dot state unchanged do not influence the exponential decay timescale leading to a fundamental difference with the $RC$-time extracted from the admittance. 

\textit{Acknowledgments.--} The authors acknowledge fruitful discussion with H. Schoeller, S. E. Nigg and M. Moskalets. This work was supported by SPP 1285, EU STREP GEOMDISS, the Ministry of Innovation NRW, the Swiss NSF and MaNEP. 



\begin{thebibliography}{24}
\bibitem{comment_renormalization} Renormalization corrections, that in our theory appear naturally, can 
              not be found in a standard perturbation expansion, where they are usually introduced by hand at
               the end of the calculation.
\bibitem{note_displacement} Assuming the circuit model of Fig.~\ref{fig_model}, the displacement current 
                     is proportional to the tunneling current~\cite{bruder94} and therefore has no influence on the 
                     $RC$ time.
\end{thebibliography}
\end{document}